# Observation of Multiple folding Pathways of β-hairpin Trpzip2 from Independent Continuous Folding Trajectories

Changjun Chen and Yi Xiao[*]
*Biomolecular Physics and Modeling Group, Department of Physics*
*Huazhong University of Science and Technology, Wuhan 430074 , Hubei, China*

**ABSTRACT**

We report 10 successfully folding events of trpzip2 by molecular dynamics simulation. It is found that the trizip2 can fold into its native state through different zipper pathways, depending on the ways of forming hydrophobic core. We also find a very fast non-zipper pathway. This indicates that there may be no inconsistencies in the current pictures of β-hairpin folding mechanisms. These pathways occur with different probabilities. "zip-out" is the most probable one. This may explain the recent experiment that the turn formation is the rate-limiting step for β-hairpin folding.

## Introduction

β-hairpin is one of the most important structural elements in proteins. Since its formation involves long-range interactions, understanding its folding mechanism will help us to understand those of proteins [1]. Up to now, different folding mechanisms of β-hairpins have been proposed [2-7]. Munoz et al. [2, 3] suggested a "zip-out" model from their experiments of the C-terminal β-hairpin of the B1 domain of protein G. This model assumes that the folding initiates at the turn and propagates toward the tails by forming cross hydrogen bonds sequentially and so that the hydrophobic cluster forms relatively later. This folding pathway was supported by later experiment and simulation evidences [8-10]. However, Munoz et al. [2] indicated that this is just the most probable way to form hairpin structure and other mechanisms could play a role too.

---

[*] Corresponding author.

They mentioned a "zip-in" like pathway with two ends first approaching each other to form a loop. Dinner et al. [5] proposed a "middle-out" model. This model suggests that the folding proceeds by forming a partial hydrophobic cluster and then the hairpin hydrogen bonds propagate outwards in both directions from the partial cluster. This kind of folding pathway was also supported by other simulation studies [7,11-13]. Zhou and co-workers [14] found that the hydrophobic core and the inter-strand hydrogen bonds could also form at the same time. Some simulations based on reduced models even suggested a "reptation" pathway [6, 15].

From above we can see that there are inconsistencies in the folding mechanism of β-hairpins. Experiments strongly support the zip-out model, while most simulations prefer the hydrophobic or zip-in model [3]. One of the reasons may be that all-atom level simulations usually didn't observe enough folding events. Another reason is that the experiments mainly observed the most probable pathway. Due to the requirement of cooperation of side chains of residues, it is difficult to simulate the complete folding of β-hairpin at all-atom level with standard molecular dynamics (MD) simulation method [11, 16]. Therefore, most simulation works above were carried out by supplementary methods [17-25], such as replica-exchange method, transition-path sampling and reduced model. However, these methods, especially those for all-atom models, usually use abnormal conditions. For example, replica-exchange method is a multi-replica and multi-temperature method and leads to high sampling efficiency. But at the same time it would also cause discontinuity in the trajectories and over-sampling in the high energy states. Therefore, few independent continuous folding trajectories of β-hairpins have been obtained.

In this letter, we report successful simulations of the folding of hairpin trpzip2 [26] by using all-atom molecular dynamics simulation method [27] at room temperature. The sequence of the trpzip2 is SWTWENGKWTWK (Fig.1a). Its native β-hairpin structure has an obvious hydrophobic core, which is composed of two aromatic side-chain pairs (Fig.1b). Since these two hydrophobic pairs are close to the turn and the tail respectively, we shall show that this can makes trpzip2 fold along different pathways and become a very ideal model for investigating the folding mechanisms of β-hairpins.

## Methods

We have done thirty-eight 100-ns simulations for the trpzip2 with AMBER PARM96 force field [28] and GB/SA implicit solvent model [29, 30] by using Tinker (See: http://dasher.wustl.edu/tinker/). All the simulations are carried out at 298K and one Atm. The time step is 1 fs. The initial structure is an extended β-strand (Fig. 1b).

For illustration of the folding processes, we select three order parameters: the root-mean-square difference (RMSD), radius of gyration of aromatic pairs ($R_g$) and inter-strand hydrogen bonds ($N_h$). The RMSD indicates the similarity between any structure and the native state and it only involves the $C^\alpha$ atoms in the backbone. The radius of gyration of aromatic pairs ($R_g$) is the value corresponding to size of the hydrophobic core.

$$R_g = R_g(2,11) + R_g(4,9), \qquad (1)$$

where $R_g(2, 11)$ is the radius of gyration of pair Trp2-Trp11, $R_g(4, 9)$ is that of pair Trp4-Trp9.

The hydrogen bond is assumed to be formed when the distance between carbonyl oxygen O and amide hydrogen H in the backbone is shorter than 3.0 Å and the angle of Donor-Hydrogen-Acceptor is larger than 120°. For simplicity we describe all the hydrogen bonds as ($i_O$-$j_H$). It represents a hydrogen bond between the carbonyl oxygen in the residue *i* and amide hydrogen in the residue *j*. For example, the hydrogen bond network of the native conformation of trpzip2 is (5-8, 8-5, 3-10, 10-3, 1-12).

## Results and Discussions

**Free energy landscape**

To check the distribution of conformations for the trpzip2 in the folding process under normal conditions, we analyze the total thirty-eight 100 ns trajectories and construct the free energy landscape for it [31]. The relative free energy between two states is calculated by

$$F_1(x) - F_2(x) = -RT \ln(P_1(x)/P_2(x)) \qquad (2)$$

where *P(x)* is the corresponding probability distribution function, *x* is any set of order parameters

[32]. We set the order parameters as the backbone root-mean-square difference (RMSD) from the native structure and radius of gyration of aromatic pairs ($R_g$). Fig.2 shows the free energy landscape is very rough. It is not easy for the short peptide to fold from unfolded states. Totally there are seven minima in the landscape. N is the native state and M1 to M6 are the local stable states.

**Folding pathways**

To investigate the folding pathways of trpzip2, we focus on the ten successful folding events from the total thirty-eight ones. A folding event is defined as the folding from RMSD larger than 3.0 Å to lower than 1.0 Å. From these ten folding events we are surprise to find that they belong to four different pathways: zip-out, zip-in, middle-out and non-zipper. In the following, we select four typical folding events to describe these pathways respectively.

**Zip-in pathway.** Fig.3 shows the RMSD, number of inter-strand hydrogen bonds, radius of gyration of aromatic pairs and distances between the inter-strand hydrogen bonding atoms via time and Fig.4 shows some key conformations in the folding pathway. The folding proceeds as follows: After quick collapse (Fig.4a-b), the peptide falls into a hairpin-like structure. In this structure, two aromatic pairs Trp2-Trp11 and Trp4-Trp9 are formed but located between two β-strands. This prohibits the formation of inter-strand hydrogen bonds. In the following (0.3-2.1ns), the trpzip2 tries to adjust the aromatic pairs toward outsides. During this process, only inner aromatic pair is broken and its gyration increases to 7 Å (Fig.3(a)). The most outside hydrogen bond (1-12) between the two end residues formed in this step (Fig.3(b)). At 2 ns, the hydrophobic core is just in the native conformation. The RMSD decreases to 1.0 Å. Then, the native inter-strand hydrogen bonds (3-10, 10-3) and (5-8, 8-5) are formed at 2.1 and 6.9ns respectively (see Fig.4d and Fig.3(b)). Finally, the "turn" region adjusts to the correct configuration (Fig.4e). These show the native hydrogen bonds form in the order: (1-12)→(3-10, 10-3)→ (5-8, 8-5). It is a "zip-in" pathway. The forming of the hydrogen bonds is trigged by forming an outer partial hydrophobic core [5, 6, 15].

**Zip-out pathway.** Fig.5-Fig.6 also illustrate a folding event. In this case, after initial collapse (Fig.6a-b), trpzip2 also forms a hairpin-like structure that is approximately a mirror of the native state like the local state M3 (Fig.6c). All the aromatic side-chains are placed on the other side of the plane of the backbone. In this structure, the native hydrogen bond (5-8) is already formed (see Fig.5(b)). At about 27 ns, the peptide extends by breaking the outer aromatic pairs but keeping the inner pairs. Fig.5(a) shows that the radius of gyration of outer aromatic pairs (Trp2-Trp11) increases suddenly over 7 Å at 27 ns while the gyration of inner aromatic pair (Trp4-Trp9) keeps lower than 5 Å. Finally at about 50 ns, trpzip2 re-collapses and the aromatic pairs aggregate at the right direction. Then, the native inter-strand hydrogen bonds form from the turn to the tail in the order: (5-8) → (8-5, 3-10, 10-3) → (1-12). In this process "turn" region, i.e. the inner aromatic pair (Trp4-Trp9) and inner hydrogen bond (5-8), is very crucial to the hairpin folding. It is clear that this folding pathway is a "zip-out" one [2, 3].

**Middle-out pathway.** Fig.7-Fig.8 illustrate another folding event. After initial fast collapsing period, trpzip2 forms a hairpin-like structure with the two hydrophobic pairs opposite to the native one as above (Fig.8a-b)), but in this case no any native hydrogen bonds are formed (see Fig.7(b)). In the following process, the trpzip2 costs about 30 ns to adjust the hydrophobic pairs by extending (Fig.8c)) and breaking both aromatic pairs (Fig.7(a)). At about 41ns, the peptide re-collapses and the two aromatic pairs form almost simultaneously. At the same time, the middle native hydrogen bonds (3-10, 8-5, 10-3) are also formed sequentially (Fig.7(b)). Then the two most outer native hydrogen bonds (5-8, 1-12) form approximately at about 47ns. In this process, the two aromatic pairs form a hydrophobic core at the middle and then initiate the formations of inter-strand hydrogen bonds propagating outwards in both directions. This just likes a "middle-out" pathway [15].

**Non-zipper pathway.** We also find a very fast folding pathway. In this case, the peptide quickly collapses into a hairpin-like structure similar to the native one. Then, the peptide adjusts its conformation and folds into the native state in less than 0.2ns. The two hydrophobic pairs and the five native hydrogen bonds are formed almost simultaneously. It is clear that this folding pathway is not a zipper one. Zhou and co-workers also suggested such a pathway based on their simulations of the C-terminal β-hairpin of protein G with a highly parallel replica exchange method [14].

Among the ten folding events, there are five "zip-out", one "zip-in", one "middle-in" and three "non-zipper" pathways. This indicates that "zip-out" is the most probable folding pathway, i.e., it is the mostly-observed pathway. This may explain the results of recent experiment and some Monte Carlo simulations that the turn formation is the rate-limiting step for β-hairpin folding. However, this does not exclude other mechanisms, although with lower probability of occurrence. It is noted that our results show that the fast non-zipper pathway also occurs with large probability. However, this non-zipper pathway may be very difficult to observe experimentally because the folding is very fast. We did not observe the "reptation" pathway. The reasons may be that we do not have enough folding events or that reptation moving is difficult to realize for real peptides since it has only been observed in reduced peptide models without side-chains.

It is noted that the frequency of observations of the zip-out versus zip-in (and middle out) is likely proportional to the relative strength of hydrogen bonds and hydrophobic interactions. The hairpin trpzip2 is not natural but designed, and have stronger hydrophobic interactions due to the two aromatic pairs. Besides, hydrophobic interactions are typically overestimated with implicit solvent models. The stronger hydrophobic interactions produced a rough landscape with many local stable states, as shown by Fig.2. Most of these local states are stabilized by hydrophobic cores. This makes trpzip2 fold more probably along zip-in or middle-out pathways. If we deduce the strength of hydrophobic interactions, e.g., mutating the inner aromatic pair and remaining the outer one (similar to beta-hairpins in nature), we would obtain a smoother landscape which makes the zip-out mechanism more probable. These indicate that the zip-out mechanism may be more pervading in nature than observed in our simulations.

**Conclusion**

In summary, we observed multiple folding pathways for the trpzip2, depending on how the two hydrophobic pairs approach to their native conformations. They include three zipper pathways and one non-zipper pathway. Our results suggest that there may be no inconsistencies in the current pictures of β-hairpin folding mechanisms. All the previously proposed folding pathways may occur but with different probabilities. The "zip-out" pathways are also initiated by the formation of a partial hydrophobic core. Our results also suggest that more independent continuous folding trajectories at normal conditions are needed to provide a complete picture of β-hairpin folding mechanisms.

*Acknowledgments:* This work is supported by the NSFC under Grant No. 30525037 and 30470412.

## Figure Captions

Fig.1   (Color online )The native conformation (a) and its initial structure (b) of trpzip2 in our simulations.

Fig.2   (Color online ) Free energy landscape for peptide trpzip2. The two order parameters are selected as root-mean-square difference (RMSD) and radius of gyration of aromatic pairs ($R_g$). The point N is corresponding to native state. The points M1 to M6 indicate the local minima.

Fig.3 The parameters versus time for the trpzip2 in a zip-in pathway: (a) RMSD(Å), inter-strand hydrogen bond number, gyration $R_g(4, 9)$ of inner aromatic pair Trp4-Trp9 and gyration $R_g(2,11)$ of outer aromatic pair Trp2-Trp11 (in Å). (b) Distances between inter-strand hydrogen bonding atoms in trpzip2 (5-8, 8-5, 3-10, 10-3, 1-12) via time in a-zip-in pathway. The bonds are sorted from "turn" to "end".

Fig.4   (Color online )Some representative structures in a zip-in pathway. The side-chains of aromatic residues Trp2, Trp4, Trp9 and Trp11 are highlighted here.

Fig.5   The same as Fig.3 but for the trpzip2 in a zip-out pathway

Fig.6   (Color online ) Some representative structures in a zip-out pathway.

Fig.7   The same as Fig.3 but for the trpzip2 in a middle-out pathway.

Fig.8   (Color online ) Some representative structures in a middle-out pathway

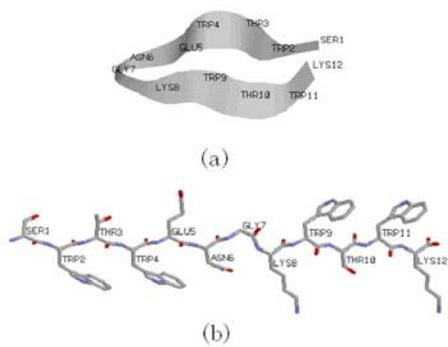

(a)

(b)

Fig. 1

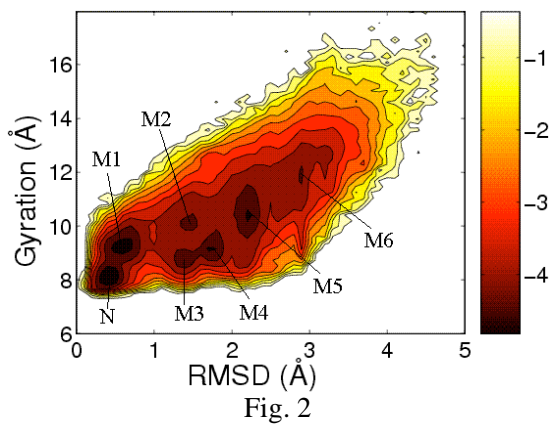

Fig. 2

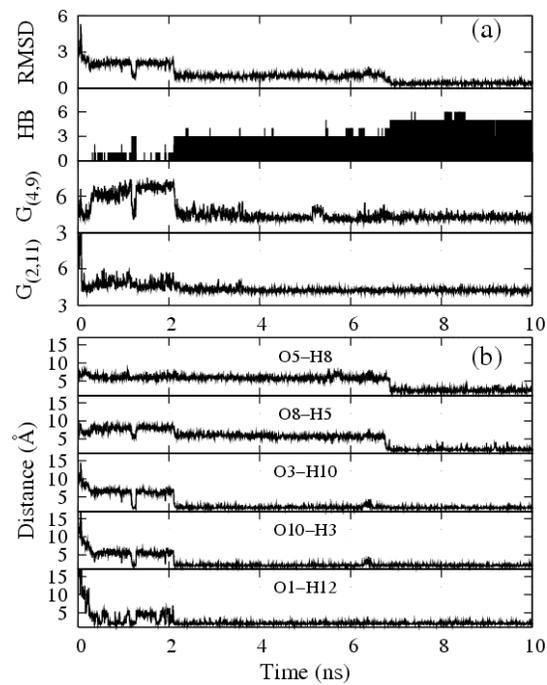

Fig. 3

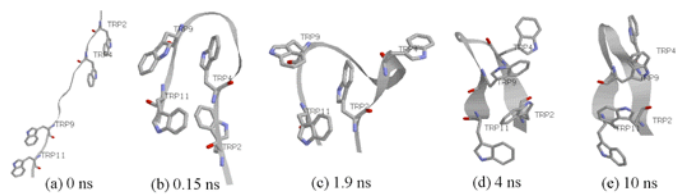

Fig. 4

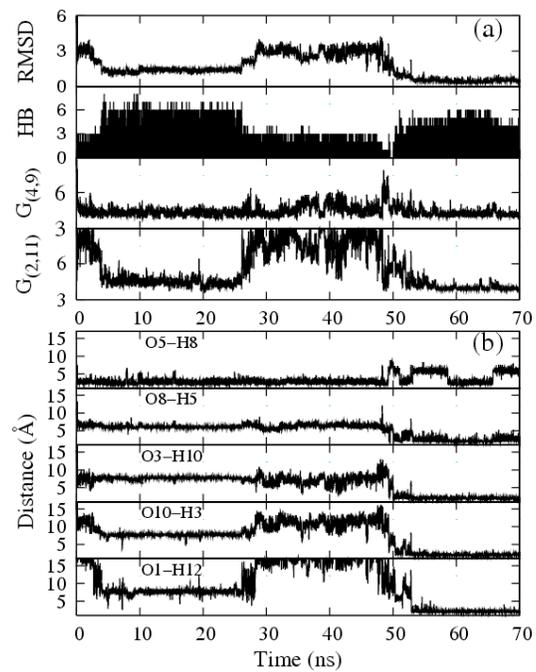

Fig. 5

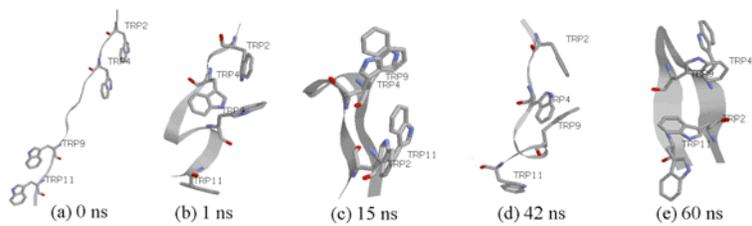

Fig. 6

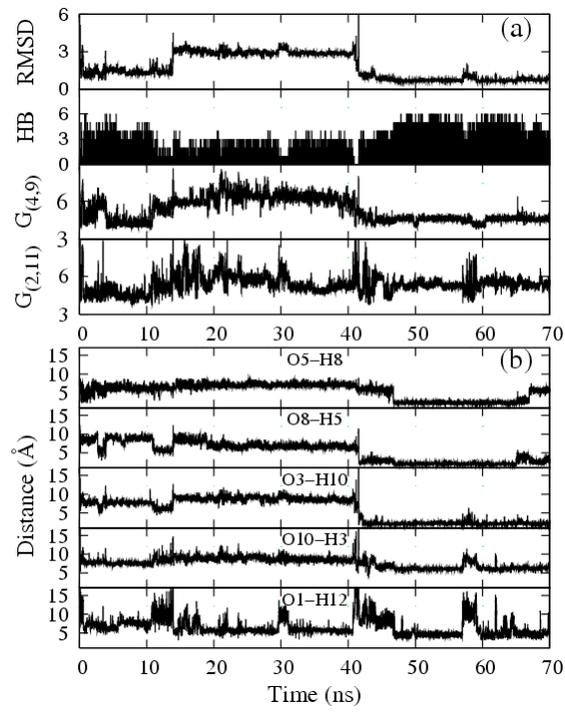

Fig.7

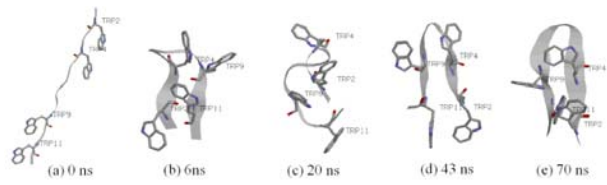

Fig.8